\documentclass[10pt, lettersize,journal]{IEEEtran}
\usepackage{amsmath,amsfonts}
\usepackage{algorithm,algorithmic}
\usepackage{array}
\usepackage[caption=false,font=normalsize,labelfont=sf,textfont=sf]{subfig}
\usepackage{textcomp}
\usepackage{stfloats}
\usepackage{url}
\usepackage{verbatim}
\usepackage{graphicx}
\usepackage{multirow}
\usepackage{makecell}
\usepackage{ulem}
\usepackage{relsize}

\usepackage{cite}
\usepackage{amssymb}
\usepackage{quantikz}

\newtheorem{definition}{Definition}
\newtheorem{example}{Example}

\newcommand{\dswap}[2]{D_{\scriptscriptstyle \text{swap}}^{\scriptscriptstyle AG}(#1, #2)}
\newcommand{\cmed}{C_{\text{mid}}}
\newcommand{\cpre}{C_{\text{pre}}}
\newcommand{\cpost}{C_{\text{post}}}
\newcommand{\slimit}{t_{\text{s}}}
\newcommand{\sdepth}{t_{\text{d}}}
\newcommand{\quekno}[1]{QUEKNO#1}
\newcommand{\revlib}[1]{RevLib#1}
\newcommand{\adacreduced}[1]{ADAC-{\textscale{0.82}{ICD}}#1}
\newcommand{\pseudorandom}[2]{$(#1, #2)$-local}

\usepackage{comment}

\begin{document}
\title{Qubit Mapping: The Adaptive Divide-and-Conquer Approach}

\author{Yunqi Huang, Xiangzhen Zhou$^*$, Fanxu Meng, and Sanjiang Li$^*$
\thanks{Yunqi Huang is with Nanjing Tech University and Centre for Quantum Software and Information (QSI), Faculty of Engineering and Information Technology, University of Technology Sydney, NSW 2007, Australia.}
\thanks{Xiangzhen Zhou is with Nanjing Tech University. (E-mail: xiangzhenzhou@njtech.edu.cn )}
\thanks{Fanxu Meng is with Nanjing Tech University. }
\thanks{Sanjiang Li is with Centre for Quantum Software and Information (QSI), Faculty of Engineering and Information Technology, University of Technology Sydney, NSW 2007, Australia. (E-mail: Sanjiang.Li@uts.edu.au)}
}

\maketitle

\begin{abstract}
The qubit mapping problem (QMP) focuses on the mapping and routing of qubits in quantum circuits so that the strict connectivity constraints imposed by near-term quantum hardware are satisfied. QMP is a pivotal task for quantum circuit compilation and its decision version is NP-complete. In this study, we present an effective approach called Adaptive Divided-And-Conqure (ADAC) to solve QMP. Our ADAC algorithm adaptively partitions circuits by leveraging subgraph isomorphism and ensuring coherence among subcircuits.
Additionally, we employ a heuristic approach to optimise the routing algorithm during circuit partitioning.
Through extensive experiments across various NISQ devices and circuit benchmarks, we demonstrate that the proposed ADAC algorithm outperforms the state-of-the-art method. Specifically, ADAC shows an improvement of nearly 50\% on the IBM Tokyo architecture.
Furthermore, ADAC exhibits an improvement of around 18\% on pseudo-realistic circuits implemented on grid-like architectures with larger qubit numbers, where the pseudo-realistic circuits are constructed based on the characteristics of widely existing realistic circuits, aiming to investigate the applicability of ADAC.
Our findings highlight the potential of ADAC in quantum circuit compilation and the deployment of practical applications on near-term quantum hardware platforms.
\end{abstract}


\section{Introduction}

\IEEEPARstart{Q}{uantum} computation, an interdisciplinary field merging quantum mechanics and computer science~\cite{feynman1982simulating}, has garnered significant attention due to its potential advantages over classical computation. Designed to exploit quantum phenomena, several quantum algorithms have demonstrated superior efficiency in solving specific problems~\cite{deutsch1992rapid,Shor94,grover1996fast,HHL}.
While these advancements underscored the promise of quantum computing, realising a universal fault-tolerant quantum hardware capable of efficiently implementing these algorithms, requires millions of qubits with low error rates and extended coherence times. Achieving such devices may require decades of research. Despite lacking error correction, the Noisy Intermediate-Scale Quantum (NISQ) era is becoming a reality~\cite{PreskillNISQ}. Intel~\cite{hsu2018intels} and Google~\cite{kelly2018preview} have already introduced quantum processors with 49 and 72 qubits, respectively. 
IBM has achieved a significant milestone with IBM Condor, a superconducting quantum processor featuring 1121 qubits based on cross-resonance gate technology~\cite{Gambetta2023ibm}.

However, utilising NISQ devices for quantum algorithms encounters challenges due to the connectivity constraints imposed by physical qubits, which pose a significant obstacle in quantum computing research. Quantum circuits, which typically allow interactions between arbitrary qubits, are constrained by the limited connectivity of NISQ devices. Therefore, it is necessary to transform quantum circuits to adhere to these connectivity constraints while maintaining functional equivalence.

SWAP gate insertion is one of the most widely used methods to achieve the aforementioned transformation, and this kind of implementation is also termed qubit mapping~\cite{Li+19-sabre}. The qubit mapping problem (QMP) involves adjusting the placement of logical qubits onto physical qubits during execution, taking into account the connectivity constraints of the qubits. In this work, we focus on addressing the challenges in QMP posed by limited connectivity constraints on NISQ devices.

\begin{figure}[htbp]
\centering

\scalebox{0.8}{
\subfloat[]{
    \includegraphics[width=0.17\textwidth]{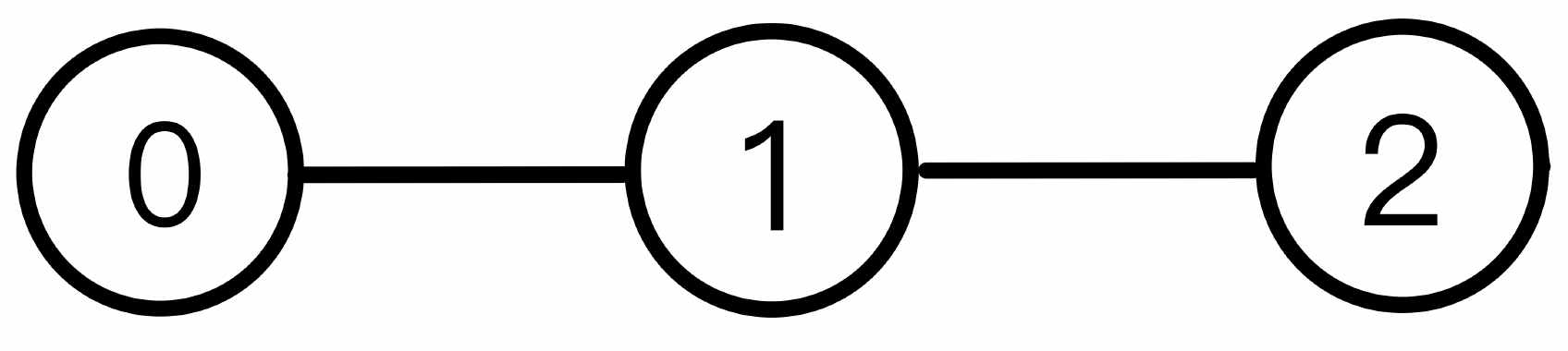}
    \label{fig:ag_line}
}
  \hfill
  }

\scalebox{0.8}{
\subfloat[]{
\begin{quantikz}
  q_0 & \ctrl{1} & \qw & \ctrl{2} &\qw \\
  q_1 & \targ{} & \ctrl{1} & \qw &\qw \\
  q_2 & \qw & \targ{} & \targ{} &\qw \\
\end{quantikz}
    \label{fig:trans_orig}
  }
  \hfill 
}  
\subfloat[]{

\scalebox{0.8}{
\begin{quantikz}
  0 & \ctrl{1} & \qw & \qw  & \ctrl{1} & \qw\\
  1 & \targ{} & \ctrl{1} & \swap{1} &\targ{} & \qw \\
  2 & \qw & \targ{} & \targX{} &\qw &\qw \\
\end{quantikz}
    \label{fig:trans_after}
}
}

  \caption{A simple example of QMP: (a) the architecture graph,   (b) the original quantum circuit, and (c) the transformed circuit.}
  \label{fig:qct}
\end{figure}

For instance, in Figure \ref{fig:qct}\subref{fig:trans_orig}, the original quantum circuit needs to be executed on a NISQ device with connectivity constraints depicted in Figure \ref{fig:qct}\subref{fig:ag_line}. Suppose we place qubit $q_0$ on 0, $q_1$ on 1 and $q_2$ on 2. The first two CNOT gates can be executed directly. However, the third CNOT gate between physical qubits 0 and 2 cannot be executed directly due to the lack of direct connectivity. To resolve this, a SWAP gate is inserted between $1$ and $2$ to reposition them, enabling the successful execution of the third CNOT gate on physical qubits 0 and 1. The transformed circuit is shown in Fig.~ \ref{fig:qct}\subref{fig:trans_after}. 

\section{Related Works}

QMP focuses on strategically placing logical qubits from a quantum circuit onto physical qubits in quantum hardware. The key objective of QMP is to minimise the insertion of auxiliary SWAP gates or reduce the depth of the transformed circuit while adhering to connectivity constraints. Importantly, it has been proven that determining whether a QMP has an optimal solution with up to $k$ SWAPs is NP-complete~\cite{Siraichi+18}.

Existing solutions for QMP can be broadly categorised into two types. The first type formulates QMP as optimisation problems and utilizes off-the-shelf tools such as Boolean satisfiability solvers~\cite{wille2019mapping}, temporal planners~\cite{ShaikP23iccad}, integer linear programming (ILP)~\cite{Almeida+19_permutation}, and SMT-solver~\cite{TanC20_iccad_optimal}. However, these solutions all face challenges due to their exponential time complexity in the worst case.

The second type comprises heuristic search methods, where circuits are transformed to satisfy the connectivity constraints by incrementally inserting auxiliary operations, such as SWAP gates. Zulehner et al.~\cite{Zulehner+18_Astar} introduced the first heuristic QMP algorithm for actual NISQ devices.  
This approach outperformed the then IBM’s solution~\cite{cross2018ibm} in terms of gate number. Note that the time complexity of the $A^\star$ algorithm ~\cite{Zulehner+18_Astar} is also exponential in the worst case. Li et al.~\cite{Li+19-sabre} proposed a polynomial QMP algorithm, called SABRE, based on a look-ahead heuristic cost function, surpassing the $A^\star$ approach~\cite{Zulehner+18_Astar} in both running time and circuit size. Several other approaches~\cite{Zhou+20_SAHS,LiZF21_fidls,zhu2020dynamic,niu2020hardware,zhu2021iterated} also utilised similar look-ahead heuristic cost functions, achieving improved performance in terms of running time and circuit size.

In addition to these approaches, another paradigm, the divide-and-conquer approach, has gained attention in addressing QMP. This strategy involves dividing the input circuit into parts and addressing each part. In fact, the $A^\star$ algorithm \cite{Zulehner+18_Astar} can be identified as a divide-and-conquer approach. This work partitioned the input circuit into layers and transformed the circuit layer by layer. Another instance of this approach is in Qiskit's StochasticSWAP method~\cite{qiskitIBMshort}, which partitioned the circuit into layers and randomly generated SWAP gates to make the following layer executable. Furthermore, Ref.~\cite{ChildsSU19-qct} also adopts a similar divide-and-conquer idea, employing various heuristic strategies for qubit placement.

Besides layer-wise partition, subgraph isomorphism has emerged as another approach within the divide-and-conquer methodology for addressing QMP.
In Ref.~\cite{siraichi+19_bmt}, the subgraph isomorphism was integrated with token swapping to solve the QMP. 
Ref.~\cite{iccad22/Wu_robust} went one step further, obtaining a better circuit partition with more similarities between consecutive sub-circuits by controlling their gates number.

Combining subgraph isomorphism and circuit partition with heuristic method,
in this paper, we introduce the adaptive divide-and-conquer (ADAC) algorithm for solving QMP. ADAC initially identifies the largest subgraph isomorphic subcircuit within the input logical circuit, ensuring that the corresponding subcircuit can be executed directly on the device without additional SWAP gates. The algorithm then determines the best initial mapping for this subcircuit, considering the efficient execution of the remaining subcircuits on both the left and right. 
During the routing process, ADAC adaptively partitions the remaining circuits and inserts the corresponding SWAP gates based on some heuristic methods.
This systematic approach effectively reduces the added SWAP gates in specific scenarios, leading to improved performance for quantum circuit execution on NISQ devices.

Our method is closely related to \cite{siraichi+19_bmt} and ~\cite{iccad22/Wu_robust}, which also divide the input circuit based on the subgraph isomorphism property of the subcircuits when processing. The proposed algorithm in \cite{siraichi+19_bmt} first divides the gates list of input circuit into maximal isomorphic sublists, and then constructs for each sublist a bounded number of embeddings and combines any two consecutive sublists by token swapping \cite{miltzow2016approximation}. The optimal transformation path is found in a dynamic programming style. 
The method proposed in \cite{iccad22/Wu_robust} divides the circuits using SMT-based check while limiting the number of edges in a sub-circuit to facilitate similarity maximisation between two successive sub-circuits.  The two methods are based on exhaustive search or SMT solver, which are not scalable and often difficult to handle circuits with 20 or above qubits.
In our ADAC, we divide the circuit \textit{adaptively} according to the number of SWAPs that need to be inserted based on a heuristic method. While dividing the circuits, we directly provide the required SWAPs and corresponding embeddings, which makes our method scalable.

Our work uses SABRE~\cite{Li+19-sabre} as the baseline method. 
SABRE is widely recognised as a state-of-the-art solution in the field, integrated into the framework of Qiskit\cite{qiskit-sdk-py}. Its effectiveness and efficiency in optimising the number of SWAP have earned it recognition as a leading solution in the field~\cite{li+23quekno}.
Compared with SABRE, our ADAC algorithm shows an advantage on IBM Tokyo architecture.
We evaluated ADAC against extensive realistic circuit benchmarks such as RevLib \cite{wille2008revlib}, QUEKNO \cite{li+23quekno}, and  the circuit library in Qiskit, where improvements were demonstrated at 68.83\%, 20.21\%, and 55.45\%, respectively.
To further assess ADAC's applicability, we designed pseudo-realistic circuits inspired by the characteristics of relevant benchmarks. Experimental results indicate that ADAC maintains approximately 18\% and 20\% advantagse as the number of qubits increases on grid-like architectures and Google Sycamore architecture. 
Additionally, we conducted experiments on Google Sycamore for real circuits possessing pseudo-realistic characteristics, such as piecewise polynomial Pauli rotation circuits and Integer comparator circuits. ADAC consistently maintained around a 30\% performance advantage regardless of the number of qubits involved.

The remainder of this paper is organised as follows. In Sec.\ref{sec:bg}, we recall some background of QMP and introduce the notations and definitions used in this paper. Sec.\ref{sec:adac} describes our ADAC algorithm. The experimental results are shown in Sec.\ref{sec:eva}. The last section concludes the paper with a discussion of limitations and directions for future research.

\section{Preliminaries and notations}

\label{sec:bg}
In this section, we briefly introduce the background and useful notations of QMP. We also introduce the definition of subgraph isomorphism and SWAP distance.

\subsection{Basics of quantum circuit}

\subsubsection{Qubit} 
Quantum computing leverages the principles of quantum mechanics to perform computations. Quantum bits, or qubits, are the basic units of information in a quantum computer. Unlike classical bits, which can be either 0 or 1, qubits can exist in a superposition of states.
A qubit is described by a quantum state vector, often denoted as $|\psi\rangle$. For a single qubit, the general state can be expressed as: $|\psi\rangle = \alpha |0\rangle + \beta |1\rangle$. Here, $|0\rangle$ and $|1\rangle$ are the basis states corresponding to classical bit values 0 and 1, respectively. The coefficients $\alpha, \beta \in \mathbb{C}$ are probability amplitudes satisfying $|\alpha|^2 + |\beta|^2 = 1$.

The ability of qubits to exist in superposition allows quantum computers to perform certain types of calculations much more efficiently than classical computers. Additionally, qubits can be entangled, leading to quantum entanglement, another quantum phenomenon that plays a crucial role in quantum information processing.

\subsubsection{Quantum gate} Quantum gates are fundamental operations in quantum computing that manipulate the state of qubits. They can be broadly categorised into single-qubit gates and multi-qubit gates. 

Single-qubit gates act on a single qubit and can change its state. One of the well-known single-qubit gates is the Hadamard gate. For a qubit in state $|0\rangle$, the Hadamard gate transforms it into the superposition state $(|0\rangle + |1\rangle) / \sqrt{2}$, and for $|1\rangle$, it becomes $(|0\rangle - |1\rangle)/\sqrt{2}$. Other examples of single-qubit gates include Pauli gates $X, Y, Z$, phase gates, and more.

Multi-qubit gates operate on more than one qubit simultaneously. The CNOT (Controlled NOT) gate is a commonly used two-qubit gate. It performs a NOT operation on the target qubit (flips its state) only if the control qubit is in the state $|1\rangle$.
Another important multi-qubit gate is the SWAP gate,
which exchanges the states of two qubits. 
When a SWAP gate is applied on qubits $q$ and $q'$, it is represented as SWAP$(q,q')$.

\begin{figure}
    \centering
    \includegraphics[width=0.25\textwidth]{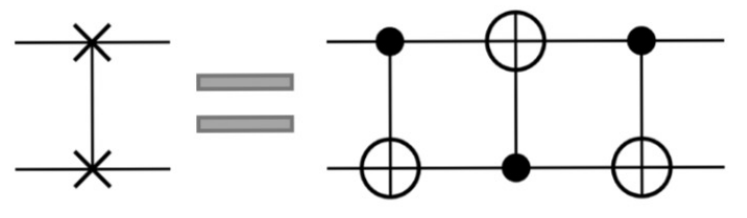}
    \caption{A decomposition of SWAP gate.}
    \label{fig:swap_dec}
\end{figure}

It's worth noting that any quantum gate operation can be decomposed into a combination of single-qubit gate and CNOT gates~\cite{BarencoElementaryGates}. This property is known as gate universality, allowing a universal set of gates to perform any quantum computation. Therefore, we only consider single-qubit and CNOT gates in this work. For the SWAP gate, it can be decomposed into three CNOT gates, shown in Figure \ref{fig:swap_dec}.

\subsubsection{Quantum circuit} 
In quantum computing, a quantum circuit is a model that represents the flow of information and operations in a quantum algorithm. A quantum circuit consists of qubits and quantum gates that manipulate these qubits' state. In this work, we use $(Q, C)$ to denote a quantum circuit, where $Q = \{q_1,\ldots,q_n\}$ is the set of qubits and $C = \{g_1, \ldots,g_m\}$ is the sequence of quantum gates. 
We use $C^{-1}$ to represent the reverse of $C$, i.e., $C^{-1} = \{g_m, \ldots,g_1\}$.

In the remainder of this paper, we assume that all gates in a quantum circuit $C$ are native gates in the target NISQ device. Without loss of generality, we assume CNOT is the only native two-qubit gate. Since the actual function of a single-qubit gate $g$ acting on qubit $q$ is not important in QMP, we simply write $g$ as $\langle q\rangle$. Moreover, a CNOT gate with control qubit $q$ and target qubit $q'$ is denoted as $ \langle q, q'\rangle$.

\begin{figure}
\centering
\subfloat[]{
    \includegraphics[width=0.2\textwidth]{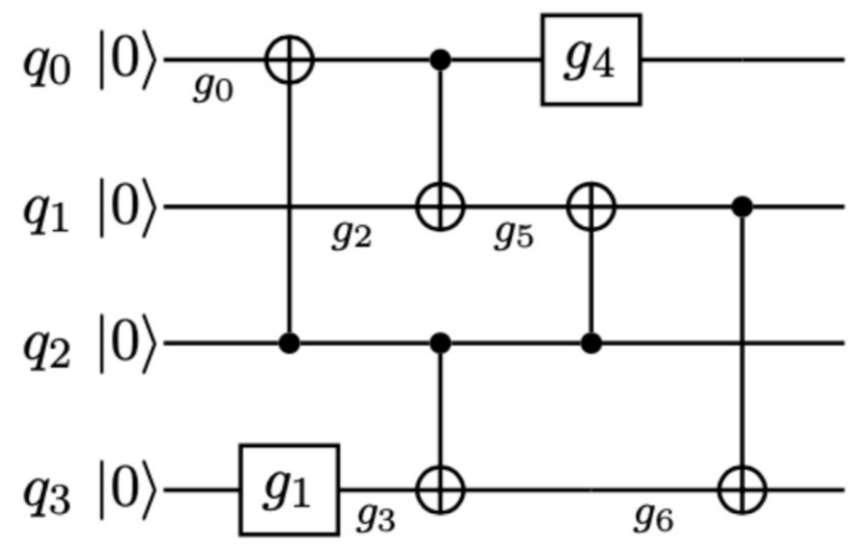}
    \label{subfig:qc}
}
\hfill
\subfloat[]{
\centering
    \includegraphics[width=0.1\textwidth]{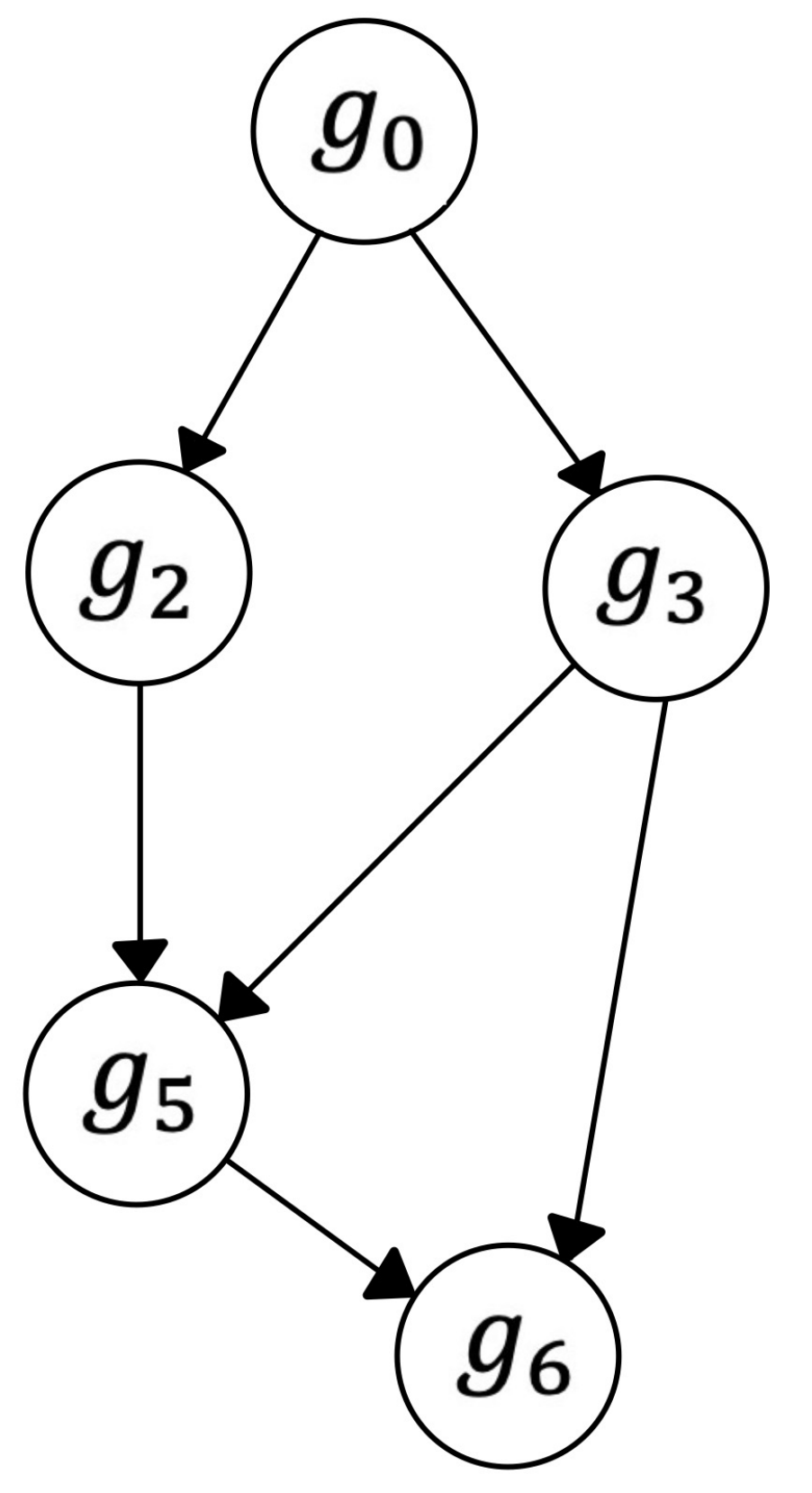}
    \label{subfig:dag}
}
\hfill
\subfloat[]{
\centering
    \includegraphics[width=0.1\textwidth]{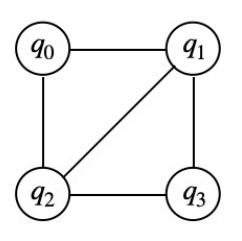}
    \label{subfig:ig}
}
\caption{A quantum circuit (a)  and its   dependency graph (b) and interaction graph (c).}
\label{fig:qc_dag}
\end{figure}

\begin{example} 
Consider the quantum circuit $(Q, C)$ shown in Figure \ref{fig:qc_dag}\subref{subfig:qc}, where $Q = \{q_0,\ldots, q_3\}, C = \{g_0 \equiv  \langle q_2, q_0 \rangle, g_1\equiv \langle q_3\rangle, g_2 \equiv  \langle q_0, q_1\rangle, g_3 \equiv  \langle q_2, q_3 \rangle, g_4\equiv \langle q_0\rangle, g_5 \equiv  \langle q_2, q_1 \rangle, g_6 \equiv  \langle q_1, q_3\rangle \}$. Then $C^{-1} = \{g_6,\ldots,g_1\}$.
\end{example}

For a quantum circuit $(Q, C)$, the dependency graph  serves as a graphical representation of the relationships between two-qubit gates. The definition is shown as follows.
\begin{definition}[Dependency Graph]
Given a quantum circuit $(Q,C)$, the corresponding dependency graph is a direct acyclic graph (DAG), where the nodes represent the two-qubit gates in $C$, and the directed edge from gate $g_i$ to $g_j$ indicates that the execution of $g_j$ depends on the execution of $g_i$.
\end{definition}
The dependencies reflect that a two-qubit gate $\langle q_i, q_j \rangle \in C$ can be executed only when all the previous two-qubit gates on $q_i$ or $q_j$ have been executed. In the dependency graph, the \textit{front layer} is defined as the set of all two-qubit gates with no incoming edges, meaning they have no dependencies on other gates within the circuit. These gates can be executed without waiting for the completion of any preceding gates. We say a gate $g \in C$ is executable if $g$ is in the front layer of $C$'s dependency graph.

\begin{example}
Figure \ref{fig:qc_dag}\subref{subfig:dag} shows the corresponding dependency graph of the circuit in Figure~\ref{fig:qc_dag}\subref{subfig:qc}. The front layer of this dependency graph is $g_0$. If $g_0$ and $g_1$ are executed, the front layer will be $g_2$ and $g_3$. 
\end{example}

For the gate sequence  $C$ in a quantum circuit, we can represent qubit interactions using a graph.  The definition can be expressed as follows:

\begin{definition}[Interaction Graph]
Given a quantum circuit $(Q,C)$, the corresponding interaction graph, denoted as $IG(C)$, is an undirected graph where each node represents a logical qubit used in the gate from $C$. If a two-qubit gate in $C$ exists that acts on a pair of logical qubits, then an edge is connecting the corresponding nodes in $IG(C)$. 
\end{definition}

\begin{example}
For the quantum circuit $(Q,C)$ in Figure \ref{fig:qc_dag}\subref{subfig:qc}, there exist five two-qubit gates $g_0 \equiv \langle q_2, q_0 \rangle,g_2 \equiv \langle q_0, q_1\rangle, g_3 \equiv \langle q_2, q_3 \rangle,g_5 \equiv \langle q_2, q_1 \rangle$ and $g_6 \equiv \langle q_1, q_3\rangle$.
The interaction graph of $C$ includes four nodes corresponding to $q_0, q_1, q_2$ and $q_3$. In addition, the edges are $(q_2, q_0), (q_0, q_1), (q_2, q_3), (q_2, q_1) $ and $(q_1, q_3)$. $IG(C)$ is shown in Figure \ref{fig:qc_dag}\subref{subfig:ig}.
\end{example}

\subsection{NISQ devices} 
\begin{figure}
    \centering
    \subfloat[IBM Tokyo]{

        \includegraphics[width=0.23\textwidth]{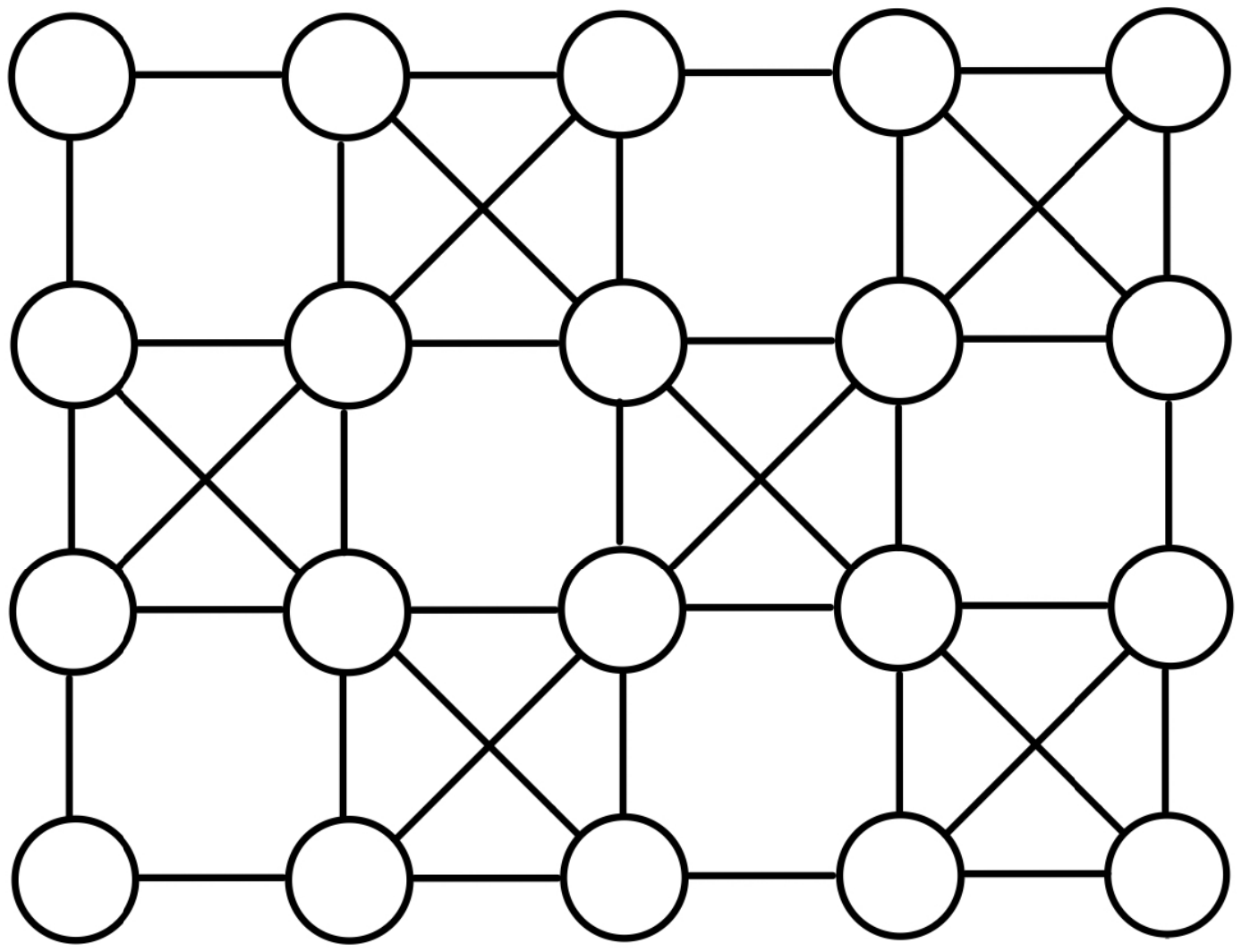}
        \label{subfig:q20}
    }
    \hfill
    \subfloat[Google Sycamore]{
        \includegraphics[width=0.2\textwidth]{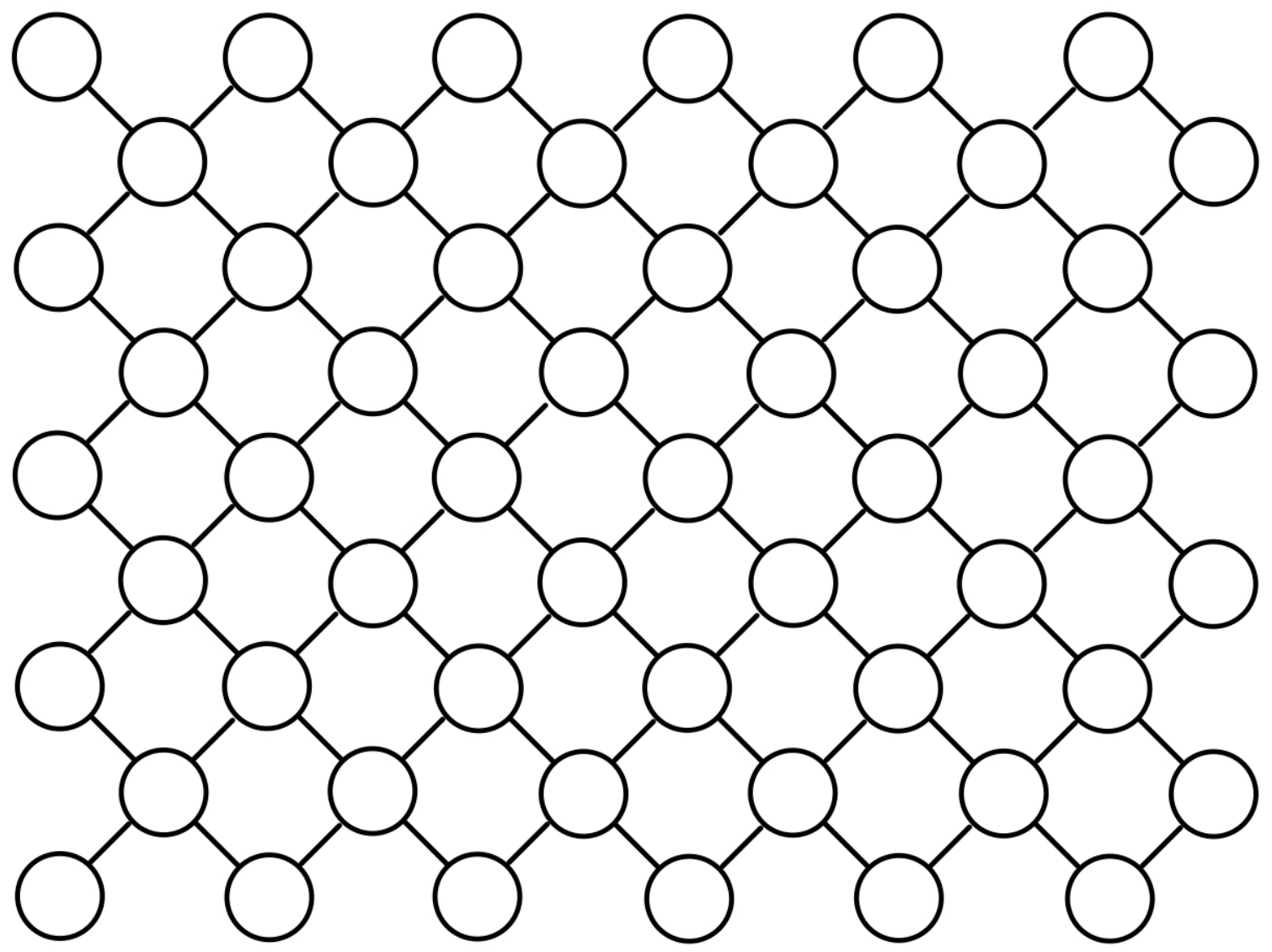}
    
        \label{subfig:syc}
    }
    \caption{Two exemplary architecture graphs.}
    \label{fig:AG}
\end{figure}

In practical quantum computing, the execution of quantum algorithms relies on NISQ (Noisy Intermediate-Scale Quantum) devices. These devices exhibit limited connectivity between physical qubits,  defined by an architecture graph $AG = (V, E)$, where $V$ is the set of physical qubits, and $E$ is the set of edges representing allowable two-qubit interactions. 
As illustrative examples, Figure \ref{fig:AG} shows two $AG$s: IBM Tokyo \subref{subfig:q20} and Google Sycamore \subref{subfig:q20}, both of which will be used in our experiments. 
 
\subsection{The qubit mapping problem}
The input of QMP is a logical circuit, denoted $LC$, without considering connectivity constraints imposed by quantum hardware.
Given a logical circuit $LC$ and a NISQ device with an architecture graph $AG$, QMP aims to compile $LC$ into a physical circuit that adheres to the connectivity constraints of the NISQ device. 
This involves placing logical qubits on physical qubits, defined by an injective function $\tau: Q \to V$ such that $\tau(q) = \tau(q')$ if and only if $q = q'$ for any $q, q' \in Q$,  where $Q$ is the set of logical qubits and $V$ is the set of physical qubits. 
Given a logical circuit $LC = (Q,C)$, the architecture graph $AG = (V, E)$ and a mapping $\tau : Q \to V$, 
if $\forall \langle q, q'\rangle \in C, (\tau(q),\tau(q')) \in E$, we say $\tau$ satisfies $C$ in $AG$, which means all the gates in $C$ could be directly executed under the mapping $\tau$ on $AG$.

\subsection{Subgraph isomorphism and SWAP distance}
\label{subsec:graph}
Subgraph isomorphism plays a crucial role in our approach to solving QMP. In this context, we consider connectivity constraints to determine whether the interaction graph of a sub-circuit can be embedded into the architecture graph of a quantum device.
We give the formal definition as follows.
\begin{definition}[Subgraph isomorphism]
Given two undirected graphs $H=(V, E)$ and $G=(V', E')$, if there exists an injective mapping function $\tau: V \to V'$ such that for every edge $(u, v) \in E$, there is an edge $(\tau(u), \tau(v))\in E'$, then $H$ is said to be subgraph isomorphic to $G$, denoted as $H \le G$. In this case, we also say $H$ is embeddable in $G$. We call $\tau$ an embedding from $H$ to $G$ denoted by $H \leq_{\tau} G$. If $H$ is not subgraph isomorphic to $G$, we denote as $H \nleq G$.
\end{definition}

For a logical circuit $LC = (Q,C)$ and an architecture graph $AG = (V,E)$, it is common to insert SWAP gates in order to modify the mapping $\tau:Q\to V$, thereby enabling the execution of CNOT gates that were initially not directly implementable. Suppose we have two mappings $\tau_1, \tau_2 : Q\to V$, we can apply a sequence of SWAP gates to transform $\tau_1$ to $\tau_2$. In this work, a SWAP gate is permitted between physical qubits $q, q' \in V$ if $(q, q')$ is an edge in $AG$.

\begin{figure}
    \centering
    \includegraphics[width=0.4\textwidth]{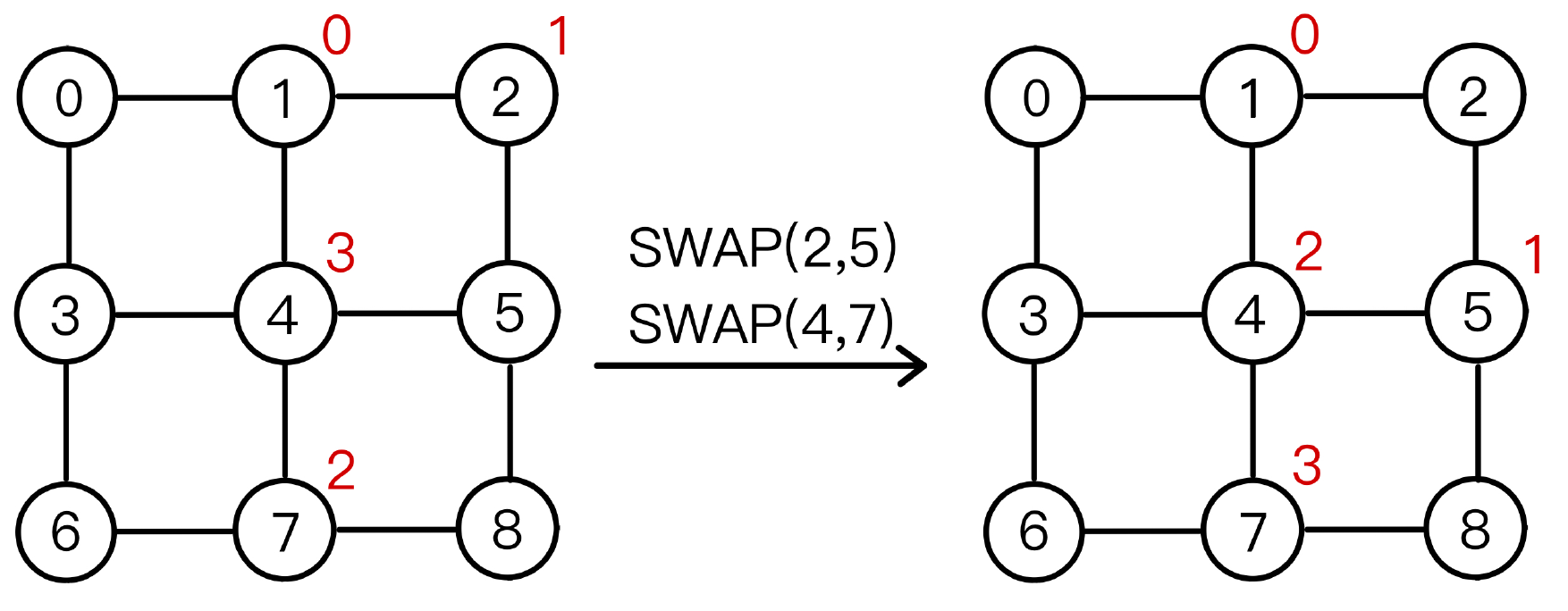}
    \caption{Applying a sequence of SWAP gates on the physical qubits.}
    \label{fig:swap_op}
\end{figure}

\begin{example}
Suppose $AG$ is shown in Figure \ref{fig:swap_op} and we have two mappings $\tau_1, \tau_2$ corresponding to the left and right sides of Figure \ref{fig:swap_op}, where $\tau_1(0) = 1, \tau_1(1) = 2, \tau_1(2) = 7, \tau_1(3) = 4$, and $\tau_2(0)=1, \tau_2(1)=5, \tau_2(2)=4, \tau_2(3) = 7$. To transform $\tau_1$ to $\tau_2$, we could apply two SWAP gates on $AG$.Specifically, we apply SWAP(2,5) to exchange the values on physical qubits 2 and 5, followed by SWAP(4,7) to exchange the values on 4 and 7. 

\end{example}

Formally speaking, for a logical circuit $LC=(Q,C)$, an architecture graph $AG = (V, E)$ and an injective mapping function $\tau : Q \to V$ such that $|Q| \le |V|$, performing a SWAP gate on edge $(u, v) \in E$ would exchange the logical qubits  mapped on $u$ and $v$, which forms an updated mapping function $\tau'$.
If no logical qubit is mapped to $u$, performing a SWAP gate on edge $(u,v)\in E$ just moves the logical qubit on $v$ to $u$. 
The procedure of updating $\tau$ is denoted as $\tau' = \tau * (u,v)$. If we obtain $\tau'$ by performing a sequence of SWAP gates represented as SWAP\_seq on $\tau$, this procedure is denoted as $\tau' = \tau * \text{SWAP\_seq}$.
Now, we know that we can transform $\tau$ into $\tau'$ by performing a series of SWAP gates. 
In this paper, two definitions of SWAP distances that measure the minimal number of SWAP gates needed to transform a mapping into another one are introduced.

\begin{definition}[SWAP distance between mappings]
Given a quantum circuit $LC = (Q,C)$, an architecture graph $AG = (V, E)$ and two mappings $\tau_1, \tau_2: Q\to V$, we define the minimal number of required SWAP gates that transform $\tau_{1}$ to $\tau_{2}$ as the SWAP distance between $\tau_{1}$ and $\tau_{2}$, which is denoted as $\dswap{\tau_{1}}{\tau_{2}}$.
\end{definition}

Defining the SWAP distance between mappings is highly intuitive. In our work, we also need to consider the number of SWAPs required to transform one mapping into the embedding of another interaction graph. This consideration is crucial in our circuit division process, as it helps maintain a slight discrepancy between successive subcircuits. The SWAP distance between a mapping and an interaction graph is defined as follows.
\begin{definition}[SWAP distance between a mapping and a graph]
Given a quantum circuit $LC = (Q,C)$ and an architecture graph $AG = (V, E)$, 
we have an embedding set $M = \{\tau' | IG(C) \le_{\tau'} AG\}$. Given a mapping $\tau : Q\to V$, we define the SWAP distance between $\tau$ and $IG(C)$ is $\dswap{\tau}{IG(C)} = \min_{\tau' \in M} \{\dswap{\tau}{\tau'}\}$.  
\end{definition}

\section{ADAC algorithm}
\label{sec:adac}

In this section, we provide a comprehensive introduction to the ADAC algorithm. The ADAC algorithm takes as inputs the quantum logical circuit $LC = (Q,C)$ and the architecture graph $AG = (V, E)$ of the target NISQ device. Our algorithm consists of the following three parts:

\begin{enumerate}
    \item Initial circuit division: The gate sequence $C$ will be divided into three parts: $\cpre, \cmed$ and $\cpost$, where $IG(\cmed) \le AG$ and the number of gates in $\cmed$ is as many as possible.
    \item Initial mapping: For $IG(\cmed)$, we find a proper embedding $\tau_{\text{ini}}$ that can execute the remaining circuit more efficiently. 
    \item Circuit division and routing: For the remaining parts, taking $\tau_{\text{ini}}$ as the initial mapping, we execute the logical circuit corresponding to $\cpost$ and $\cpre^{-1}$ using our ADAC routing algorithm, respectively.
\end{enumerate}

There are also some hyperparameters whose values will be determined in the experiment. They are:
\begin{itemize}
    \item $\slimit$: SWAP count threshold to be used in Alg.~\ref{Alg:bfsh} to limit the SWAP distance between a given mapping and interaction graph.  
    \item $\sdepth$: look-ahead search depth to be used in Alg.~\ref{Alg:find_swap_h} to balance the performance and efficiency.
\end{itemize}

The main algorithm of ADAC is shown in Alg.~\ref{alg:main}.

\begin{algorithm}
\caption{ADAC}
\label{alg:main}
\begin{algorithmic}[1]
\REQUIRE a logical circuit $LC = (Q,C)$, and an architecture graph $AG$
\ENSURE a physical circuit that can be executed directly on $AG$
\STATE $\cmed \leftarrow \text{MaxSubgraphDivision}(LC, AG)$;
\STATE Construct $LC_{l} = (Q, C_{\text{pre}}^{-1}), LC_{r} = (Q, C_{\text{post}})$ according to $\cmed$;
\STATE Find an embedding $\tau_{\text{ini}}$ of $IG(\cmed)$ according to the procedure in Sec.~\ref{sec:im};
\STATE $(C_{r}, S_{r}) \leftarrow \text{ADACRouting}(LC_{r}, \tau_{\text{ini}}, AG)$;
\STATE $(C_{l}, S_{l})\leftarrow \text{ADACRouting}(LC_{l}, \tau_{\text{ini}}, AG)$;
\STATE Use $C_{l},S_{l},\cmed, C_{r}, S_{r}$ to construct the output physical circuit.
\end{algorithmic}
\end{algorithm}

The following subsections detail the above three parts and the algorithms used in Alg.~\ref{alg:main}.

\subsection{Initial circuit division}
\label{sec:icd}
The ADAC algorithm begins by dividing the gate sequence $C$ of input logical circuit, $LC = (Q,C)$, into 3 parts, $\cpre$, $\cmed$ and $\cpost$. Note that $\cpre$ and $\cpost$ correspond to subsequences on the left and right of $\cmed$, respectively.The focus initially is on deriving $\cmed$, which serves as the starting point for the subsequent routing subroutine detailed in Sec.~\ref{sec:CDR}.
The goal of ADAC is to identify the largest subgraph isomorphic subsequence of $C$ in order to create $\cmed$. To achieve this, a greedy search algorithm is employed. 
The detailed procedure for this search is outlined in Alg.~\ref{Alg:compute_gate}.

Alg. \ref{Alg:compute_gate} takes as input the logical circuit $LC=(Q, C)$ and the architecture graph $AG$. Starting from each gate in $C$, we greedily construct a subsequence $C_{\text{cur}}$ satisfying $IG(C_{\text{cur}}) \le AG$ (corresponding to the `while' loop between Line 2 and 17). It's worth noting that the judgment for subgraph isomorphic in Line 7 is accomplished by the VF2 algorithm~\cite{Cordella+04-vf2} embedded in Rustworkx~\cite{treinish2021rustworkx}, a Python library for Graph theory. Finally, the $C_{\text{cur}}$ with maximal gate number will be selected as the initial circuit division $\cmed$ (Lines 14 to 16).

\begin{algorithm}
\caption{MaxSubgraphDivision}
\label{Alg:compute_gate}
\begin{algorithmic}[1]
\REQUIRE a logical circuit $LC=(Q, C)$ and an architecture graph $AG$
\ENSURE a new gate sequence $\cmed$ that contains the maximal subgraph isomorphic part in $C$
\STATE Initialize an empty gate sequence $\cmed$;
\WHILE{$|C| > 0$}
\STATE Initialize a gate sequence $C_{\text{cur}}$ with the first gate in $C$, and remove that gate from $LC$;
\STATE $LC_{\text{copy}} \leftarrow LC$;
\FOR{each gate $g$ in front layer of $LC_{\text{copy}}$}
\STATE Initialize a gate sequence $C_{\text{new}}$ through adding $g$ to the end of $C_{\text{cur}}$;
\IF{$IG(C_{\text{new}}) \le AG$} 
\STATE $C_{\text{cur}} \leftarrow C_{\text{new}}$, and remove $g$ from $LC_{\text{copy}}$;
\ENDIF
\IF{the front layer of $LC_{\text{copy}}$ remains unchanged after a full traversal} 
\STATE \textbf{break}
\ENDIF
\ENDFOR
\IF{$|C_{\text{cur}}| > |\cmed|$} 
\STATE $\cmed \leftarrow C_{\text{cur}}$;
\ENDIF
\ENDWHILE
\RETURN $\cmed$
\end{algorithmic}
\end{algorithm}

\subsection{Initial mapping of $\cmed$}
\label{sec:im}
Unlike the traditional qubit mapping algorithm in which the initial mapping subroutine aims to provide an optimal mapping from logical to physical qubits for the starting point, i.e., the leftmost part of the input circuit, 
ADAC employs a different approach. In ADAC, the initial mapping subroutine is invoked to ensure that all gates in $\cmed$ can be executed
since $\cmed$ contains the largest number of gates. Then, the routing algorithm will start from $\cmed$ and go to both sides. Since there can be multiple embeddings of $IG(\cmed)$, this initial mapping subroutine aims to select an embedding that minimises the SWAP cost during the routing process detailed in Alg.~\ref{sec:CDR}.
It equals to find a mapping $\tau_{\text{mid}}$ such that
\begin{equation}
\tau_{\text{mid}} = \mathop {\text{argmin}}\limits_{\tau \in \{\tau' | IG(\cmed) \le_{\tau'} AG\}}(|S_\text{pre}|+|S_\text{post}|),
\end{equation}
where $S_\text{pre}$ and $S_\text{post}$ are the output SWAP sequences of the routing subroutine to be explained in Alg.~\ref{Alg:compute_part_swap} with respect to $(Q,\cpre^{-1})$ and $(Q,\cpost)$ being the input circuits and $\tau_{\text{mid}}$ the initial mappings.

Finding the optimal initial mapping $\tau_{\text{mid}}$ is hard. To overcome this hurdle, we turn to find an approximate solution 
via the reverse traversal technique introduced in SABRE \cite{Li+19-sabre}. To be specific, the initial qubit mapping process in SABRE involves multiple forward and backward runs, in which the previous final mapping serves as the current initial mapping, on the input circuit. This reverse traversal technique is applied iteratively for a specific number of times.

In the ADAC implementation, the reverse traversal process is iterated $t_{\text{ini}}$ times. During this process, the beginning and final mappings of $C$, denoted by $\tau_{\text{beg}}$ and $\tau_{\text{fin}}$ in the last round, will be recorded. Following this, we select an embedding from the interaction graph of $\cmed$ to the architecture graph as the output initial mapping $\tau_{\text{ini}}$ such that
\begin{equation}
\label{eq:taudis}
\tau_{\text{ini}} = \mathop {\text{argmin}}\limits_{\tau \in \{\tau' | IG(\cmed) \le_{\tau'} AG\}} \sum_{q \in Q} \text{dist}[\tau(q)][\tau''(q)],
\end{equation}
where $\text{dist}[a][b]$ is the Euclidean distance of $a,b$ in $AG$, $\tau'' = \tau_{\text{beg}}$ if $|\cpre| > |\cpost|$, otherwise $\tau'' = \tau_{\text{fin}}$.

In the actual implementation, the above construction process for $\tau_{\text{ini}}$ is achieved using a modified version of VF2. 
This modified VF2 algorithm utilises a depth-first search strategy with candidate pruning. In the candidate generation phase, the computation of Eq.~\ref{eq:taudis} with input embeddings has been incorporated, leading to the selection of the embedding with the minimal distance as the final output.

\subsection{Circuit division and routing}
\label{sec:CDR}
After determining the maximum subgraph isomorphic subsequence $\cmed$ and the corresponding initial mapping $\tau_{\text{ini}}$, the remaining subcircuits $LC_l = (Q, \cpre^{-1})$ and $LC_{r} = (Q, \cpost)$ need to be divided and routing. Alg. \ref{Alg:compute_part_swap} (Adaptive routing algorithm) shows the division and routing process. 

Alg. \ref{Alg:compute_part_swap} takes as input the logical circuit $LC = (Q,C)$, the initial mapping $\tau_{\text{ini}}$, and the architecture graph $AG$. 
When given the current mapping $\tau_{\text{cur}}$, 
the subsequence $C_{\text{del}}$ will be identified adaptively satisfying $IG(C_{\text{del}}) \le AG$ and $\dswap{\tau_{\text{cur}}}{IG(C_{\text{del}})} \le \slimit$, 
where $\slimit$ is the SWAP count threshold. Once the divided circuit $C_{\text{del}}$ is identified, the corresponding SWAP sequence $S_{\text{add}}$, which transforms $\tau_{\text{cur}}$ to an embedding of $IG(C_{\text{del}})$ is also determined.These values are recorded, and the associated quantities are updated (Lines 5 to 7). Finally, after the entire circuit is divided, we obtain the circuit partitions $LC_{l}$ and $LC_{r}$ along with their corresponding SWAP sequences.

\begin{algorithm}
\caption{AdaptiveRouting}
\label{Alg:compute_part_swap}
\begin{algorithmic}[1]
\REQUIRE a logical circuit $LC=(Q, C)$, initial mapping $\tau_{\text{ini}}$, and an architecture graph $AG$.
\ENSURE the required SWAP sequences $S$ to execute $LC$, and the gate sequence divisions $C_{\text{div}}$.
\STATE Initialize an empty SWAP sequence $S$;
\STATE $\tau_{\text{cur}} \leftarrow \tau_{\text{ini}}$;
\WHILE{$|C| > 0$}
\STATE $(C_{\text{del}}, S_{\text{add}}) \leftarrow \text{HeuristicDAC}(LC, \tau_{cur}, AG, \sdepth)$;
\STATE $\tau_{\text{cur}} \leftarrow \tau_{\text{cur}} * S_{\text{add}}$;
\STATE Remove $C_{\text{del}}$ from $C$;
\STATE Add $S_{\text{add}}$ to  $S$ and add $C_{\text{del}}$ to $C_{\text{div}}$;
\ENDWHILE
\RETURN $S, C_{\text{div}}$
\end{algorithmic}
\end{algorithm}

\begin{flushleft}
\begin{algorithm}
\caption{HeuristicDAC}
\label{Alg:find_swap_h}
\begin{algorithmic}[1]
\REQUIRE a logical circuit $LC=(Q, C)$, current mapping $\tau_{\text{cur}}$, an architecture graph $AG$,  and the look-ahead search depth $\sdepth$.
\ENSURE a gate division $C_{\text{cur}}$ and a SWAP sequence $S_{\text{cur}}$ in current depth.
\STATE Initializ an empty gate list $C_{\text{cur}}$, $score_\text{best} \leftarrow 0$;
\WHILE{at least 1 gate in the front layer of $LC$ has not been traversed}
\STATE $g \leftarrow$ the untraversed gate in the front layer of $LC$ with minimal distance under $\tau_{\text{cur}}$;
\STATE Add $g$ to the end of $C_{\text{cur}}$, and remove $g$ from $LC$;
\STATE  $S_\text{cur} \leftarrow$ HeuristicSWAPs($C_{\text{cur}}, \tau_\text{cur}, AG, \slimit$);
\IF{$S_\text{cur}$ is not None}
\STATE $\tau_\text{next} \leftarrow \tau_\text{cur} * S_\text{cur}$;
\IF{$\sdepth > 1$ and $|C| > 0$}
\STATE $(C_{\text{next}}, S_\text{next}) \leftarrow$ HeuristicDAC($LC, \tau_\text{next}, AG, \sdepth-1 $);
\ELSE
\STATE Initialize empty sequences $C_{\text{next}}$ and $S_\text{next}$;
\ENDIF
\STATE Calculate $score_\text{cur}$ according to Eq.\ref{eq:score};
\IF{$score_\text{cur} > score_\text{best}$}
\STATE $score_\text{best} \leftarrow score_\text{cur}$;
\STATE $C_\text{best} \leftarrow C_\text{cur}$;
\STATE $S_\text{best} \leftarrow S_\text{cur}$;
\ENDIF
\ELSE
\STATE Remove $g$ from $C_{\text{cur}}$ and add $g$ back to $LC$;
\ENDIF
\ENDWHILE
\RETURN $C_\text{best}, S_\text{best}$
\end{algorithmic}
\end{algorithm}
\end{flushleft}

In the process of Alg. \ref{Alg:compute_part_swap}, a crucial step involves invoking Alg.~\ref{Alg:find_swap_h}. This algorithm is responsible for performing the circuit division and determining the corresponding SWAP sequence required to transform the current mapping $\tau_{\text{cur}}$ into an embedding of the subgraph isomorphic subsequence of the input circuit.

To find the best SWAP sequence and circuit division, Alg.~\ref{Alg:find_swap_h} greedily constructs the subsequence $C_{\text{cur}}$ that satisfies $\dswap{\tau_{\text{cur}}}{IG(C_{\text{cur}})} \le \slimit$, ensuring the consitency between $\tau_{\text{cur}}$ and the embedding of $IG(C_{\text{cur}})$. While satisfying these conditions, a heuristic score is used in Line 13 to evaluate and select the optimal solution. The definition is shown as follows:
\begin{equation}
\label{eq:score}
score_{\text{cur}} = \frac{|C_{\text{cur}}| + |C_{\text{next}}|}{|S_{\text{cur}}| + |S_{\text{next}}|}, 
\end{equation}
where $C_{\text{cur}}, S_{\text{cur}}$ are the current circuit division and corresponding SWAP sequence, and $C_{\text{next}}, S_{\text{next}}$ are the next best circuit division and SWAP sequence. $C_{\text{next}}$ and $S_{\text{next}}$ are obtained by recursively invoking Alg.~\ref{Alg:find_swap_h} according to the look-ahead search depth $\sdepth$. Finally, Alg.~\ref{Alg:find_swap_h} will return the circuit division $C_{\text{best}}$ and corresponding SWAP sequence $S_{\text{best}}$ with best heuristic score.

In Alg.~\ref{Alg:find_swap_h}, when given the current mapping $\tau_{\text{cur}}$ and subsequence $C_{\text{cur}}$,the objective is to verify if $\dswap{\tau_{\text{cur}}}{IG(C_{\text{cur}})} \le \slimit$. This verification ensures that transitioning from $\tau_{\text{cur}}$ to an embedding of $IG(C_{\text{cur}})$ can be achieved within the specified SWAP count threshold $\slimit$. To achieve this goal, we propose a heuristic method in Alg.~\ref{Alg:bfsh}. 

\begin{algorithm}
\caption{HeuristicSWAPs}
\label{Alg:bfsh}
\begin{algorithmic}[1]
\REQUIRE a gate sequence $C_{\text{cur}}$, current mapping $\tau_{\text{cur}}$, and an architecture graph $AG$.
\ENSURE a SWAP sequence $S_{\text{pick}}$ satisfying $|S_{\text{pick}}| \le \slimit$ and $IG(C_{\text{cur}}) \le_{\tau * S_{\text{pick}}} AG$
\STATE Initialize a solution set $\mathcal{S}$ with an empty SWAP sequence;
\WHILE{$|S_\text{pick}| < \slimit$}
\STATE $S_{\text{pick}} \leftarrow \mathop{\arg\min}_{S \in \mathcal{S}}\text{Cost}(S)$;
\STATE $\mathcal{S} \leftarrow \mathcal{S}/\{ S_{\text{pick}} \}$;
\IF{$\text{Cost}(S_{\text{pick}}) = 0$} 
\RETURN $S_\text{pick}$
\ENDIF
\FOR{each associated edge $(u, v)$ in $AG$}
\STATE Initialize a new SWAP sequence $S_\text{new}$ by adding $\text{SWAP}(u, v)$ to the end of $S_\text{pick}$;
\STATE $\mathcal{S} \leftarrow \mathcal{S} \cup  \{ S_{\text{new}} \}$;
\ENDFOR
\ENDWHILE
\RETURN None
\end{algorithmic}
\end{algorithm}

Alg. \ref{Alg:bfsh} uses a solution set $\mathcal{S}$ to store potential SWAP sequences during the process. In the While-Loop, the solution with the minimal cost is selected and denoted as $S_{\text{pick}}$. The cost of a solution is calculated based on the following criteria:. 
$$
\text{Cost}(S) = \Sigma_{g=\langle q,q'\rangle \in C_{\text{cur}}} \text{dist}[\tau'(q)][\tau'(q')],
$$
where $\tau' = \tau_{\text{cur}} * S$, $S$ is a SWAP sequence in $\mathcal{S}$, and $C_{\text{cur}}$ is the gate sequence of input. 
If the $\text{Cost}(S_{\text{pick}}) = 0$,which indicates that $\tau_{\text{cur}} * S_{\text{pick}}$ forms an embedding of $IG(C_{\text{cur}})$, Alg.~\ref{Alg:bfsh} immediately returns $S_{\text{pick}}$ as the final result. If the cost is non-zero, the algorithm proceeds by appending a possible SWAP to $S_{\text{pick}}$ to explore new potential solutions.
During this process, only SWAPs associated with occupied vertices in $\tau_{\text{cur}}$ are considered (as specified in Line 8).
Therefore, in Lines 8 to 10, we add several solutions with any possible SWAP into $\mathcal{S}$. Finally, if the number of SWAP in $S_{\text{pick}}$ exceeds $\slimit$, the loop ends with returning None result.

By utilising the routing algorithm, our ADAC algorithm systematically processes the quantum logical circuit, taking into account the distinctive characteristics of each $C_{\text{div}}$ and their influence on the overall mapping.

\section{Evaluations}
\label{sec:eva}
In this section, we outline our experimental design and present the results obtained by comparing our approach with the SABRE algorithm~\cite{Li+19-sabre}\footnote{The Qiskit-embedded version of SABRE is used here.}.
SABRE is widely recognised as a state-of-the-art algorithm.  
It is worth noting that the SABRE program used here is not exactly the same as the original article~\cite{Li+19-sabre}, but the one that has been improved and integrated into Qiskit, which we addressed SABRE-Qiskit henceforth. Particularly for the benchmarks that contain numerous subgraph isomorphic components, Ref.~\cite{li+23quekno} shows that SABRE-Qiskit demonstrates superior performance compared to other leading QMP algorithms~\cite{Sivarajah20qst_tket,Zhou+20_SAHS,MCTS_QCT}.

Hence, we select SABRE-Qiskit (SABRE for short) as the baseline for evaluating our algorithm.
The tested architecture graphs are IBM Tokyo, Google Sycamore, and grid-like architectures (cf. Fig.~\ref{fig:AG}). The benchmarks being used cover both realistic and random circuits. The realistic circuits include 158 reversible classical circuits from \revlib~\cite{wille2008revlib} with qubit numbers ranging from 5 to 15 and 12 quantum circuits (e.g. Grover operator, phase oracle) extracted from the circuit library in Qiskit with qubit numbers ranging from 5 to 20. The random circuits are either extracted from \quekno{}~\cite{li+23quekno} or construed by the rules to be elaborated in Sec.~\ref{sec:adacfactor}. The improvement percentage used throughout the whole section is defined as
\begin{equation}
\rho =\ \frac{n_{\text{com}} - n_{\text{our}}}{n_{\text{com}}},
\end{equation}
where $n_{\text{our}}$ and $n_{\text{com}}$ are, respectively, the number of SWAP gates inserted by our and the compared algorithm.
Besides, we use Python as our programming language and IBM Qiskit~\cite{qiskitIBMshort} (version 1.0.2) as an auxiliary environment. For the hyperparameters in ADAC, we empirically set $t_{\text{ini}} = 3, \slimit =12$, $\sdepth = 2$. All experiments were conducted on a MacBook Pro featuring a 2.3 GHz Intel Core i5 processor and 16 GB memory.

\subsection{Effectiveness for initial circuit division}
Recall that ICD (initial circuit division, cf. Sec.~\ref{sec:icd}) is the first step of ADAC aimed at constructing a gate sequence $\cmed$ to serve as the starting point for subsequent subroutines. This process involves identifying the maximal subgraph isomorphic division within the input logical circuit. To demonstrate its efficacy, we replace the original implementation of ICD with a straightforward one that simply uses the maximum subgraph isomorphic division at the beginning of the circuit to construct $\cmed$ and denote the degraded algorithm \adacreduced{}.

Then, experiments are done to compare the performance of the original ADAC algorithm with \adacreduced{}. Specifically, the tested circuits are  
picked from \revlib{} and the architecture graph (AG) is IBM Tokyo. The results have been recorded in Table~\ref{tb:revLibq20}. It can be seen that the ADAC algorithm exhibits an average improvement of 9.84\%
over \adacreduced{}, thereby highlighting the effectiveness of the IDC subroutine.

\begin{table*}
    \centering
    \caption{The detailed experimental results of \adacreduced{} on IBM Tokyo.}
    \begin{tabular}{|c|c|c|c|c|c|}
        \hline
        \multirow{2}{*}{Circuit Name} & \multirow{2}{*}{\# Qubits} & \multirow{2}{*}{\# CNOT}  & \multirow{2}{*}{\makecell{\# Added SWAPs  \\ \adacreduced{}}} & \multirow{2}{*}{\makecell{\# Added SWAPs  \\ ADAC}}  &   \multirow{2}{*}{Improvement } \\
        & & & & & \\
        \hline
        4mod5-v1\_22 & 5 & 11 & 0 & 0  & 0  \\
        mod5mils\_65 & 5& 16 & 0 & 0  & 0  \\
        alu-v3\_34& 5 & 24 & 2  & 2  & 0 \\
        4mod5-bdd\_287 & 7 & 31 & 2  & 2 & 0  \\
        one-two-three-v0\_98 & 5 & 65 & 5 & 5  & 0 \\ 
        ising\_model\_10 & 10 & 90 & 0  & 0 & 0  \\
        ising\_model\_13 & 13 & 120 & 0  & 0 & 0 \\
        ex3\_229 & 6 & 175 & 6 &  5  & 16.67\%  \\
        alu-v2\_30 & 6 & 223 & 16 & 14  & 12.5\% \\
        con1\_216 & 9 & 415 &  41 & 25 & 39.02\% \\
        cm42a\_207 & 14 & 771 & 49 & 44 & 10.2\%  \\
        sym6\_145 & 7 & 1701 &  124 & 111  & 10.48\% \\
        hwb6\_56 & 7 & 2952  & 276 & 259  & 6.16\% \\
        ham15\_107 & 15 & 3858  & 319 & 279 & 12.54\% \\
        sym9\_148 & 10 & 9408 & 311 & 299  & 3.86\% \\
        urf2\_277 & 8 & 10066 & 1891 & 1682  & 11.05\% \\
        max46\_240 & 10 & 11844 &1270 & 1179 & 7.17\% \\
        sym9\_193 & 11& 15232  & 1226 & 1087  & 11.34\%\\
        \hline
        Sum & - & 57002  & 5538 & 4993 & 9.84\%\\
        \hline
    \end{tabular}
    
    \label{tb:revLibq20}
\end{table*}

\subsection{Results on IBM Tokyo}

Although it is offline, IBM Tokyo (cf. Fig.~\ref{fig:AG}\subref{subfig:q20}) is a significant baseline architecture due to its wide usage in experiment parts among various qubit mapping papers, e.g., Ref.~\cite{Li+19-sabre,LiZF21_fidls,TanC20_iccad_optimal,siraichi+19_bmt,Zhou+20_SAHS}. Therefore, for IBM Tokyo, thorough experiments have been done both on realistic and random circuit benchmarks (cf. Table.~\ref{tb:q20total} for overall results).

\begin{figure}
\centering
\includegraphics[width=0.4\textwidth]{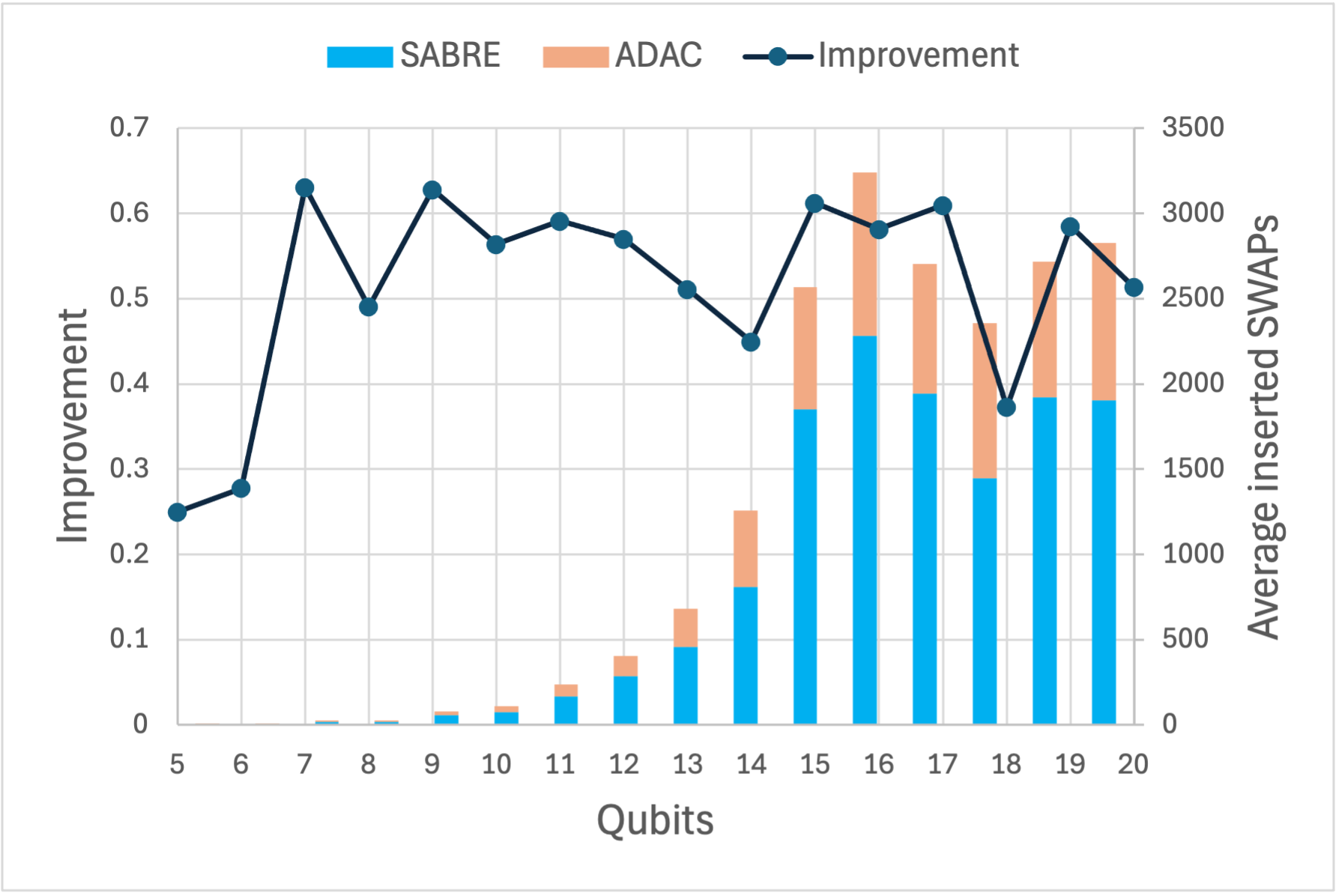}
\caption{The average improvement of ADAC over SABRE (left $y$-axis) and the average number of inserted SWAPs of SABRE and ADAC (right $y$-axis) for Qiskit circuits conducted on IBM Tokyo, where the left and right $y$-axises denote, respectively, the improvement ratio and the number of inserted SWAPs, both are averaged over circuits with the same number of qubits.}
\label{fig:realc}
\end{figure}

For the realistic circuits, the ADAC outperforms SABRE by a large margin, over 68\% in \revlib{} and 55\% in Qiskit. Additionally, Fig.~\ref{fig:realc} demonstrates ADAC's improvement variances in terms of qubit numbers for circuits in Qiskit. The results indicate that ADAC exhibits a 50\% or higher improvement most of the time. For random circuits extracted from \quekno{} dataset consisting of 200 quantum circuits, ADAC consistently shows nearly a 20\% improvement compared to SABRE. We also show the results on different range of number of CNOT gates in Table~\ref{tb:q20_cnot_range}. In Table~\ref{tb:q20_cnot_range}, we find that as the number of gates increases, our method performs better.

\begin{table}
    \caption{The overall experimental results on IBM Tokyo compared to SABRE.}
    \centering
    \begin{tabular}{|c|c|c|}
    \hline
    Circuits Name & \# Circuits & Improvement  \\
    \hline
    RevLib  & 156 & 68.83\% \\
    QUEKNO & 200 & 20.21\% \\
    Qiskit & 152 & 55.45\% \\
    \hline
    \end{tabular}

    \label{tb:q20total}
\end{table} 

\begin{table}[!ht]
    \caption{The average inserted SWAPs for SABRE and ADAC in different range of CNOT gates on IBM Tokyo}
    \centering
    \begin{tabular}{|c|c|c|c|}
    \hline
        \# CNOT  range & \# Circuits & SABRE & ADAC  \\ \hline
        $\le$1000 & 374 & 27.78 & 17.77 \\ 
        (1000, 5000] & 45 & 415.9& 131.7  \\ 
        (5000, 15000] & 25 & 1621 & 519.7  \\ 
        (15000, 30000] & 15 & 4025 & 1268 \\ 
        $>$30000 & 43 & 7007 & 2525 \\ \hline
    \end{tabular}
    \label{tb:q20_cnot_range}
\end{table}

The results in Table~\ref{tb:q20total} show that different circuits exhibit different levels of improvement. In light of these results, we carried out a thorough analysis of the specific circuit characteristics. Our research indicates that our approach yields greater performance for circuits containing multiple small-size sub-circuits. We will delve into this in detail in the next subsection.

\subsection{Results on large AGs}
\label{sec:adacfactor}

\begin{figure}
    \centering
    \includegraphics[width = 0.48\textwidth]{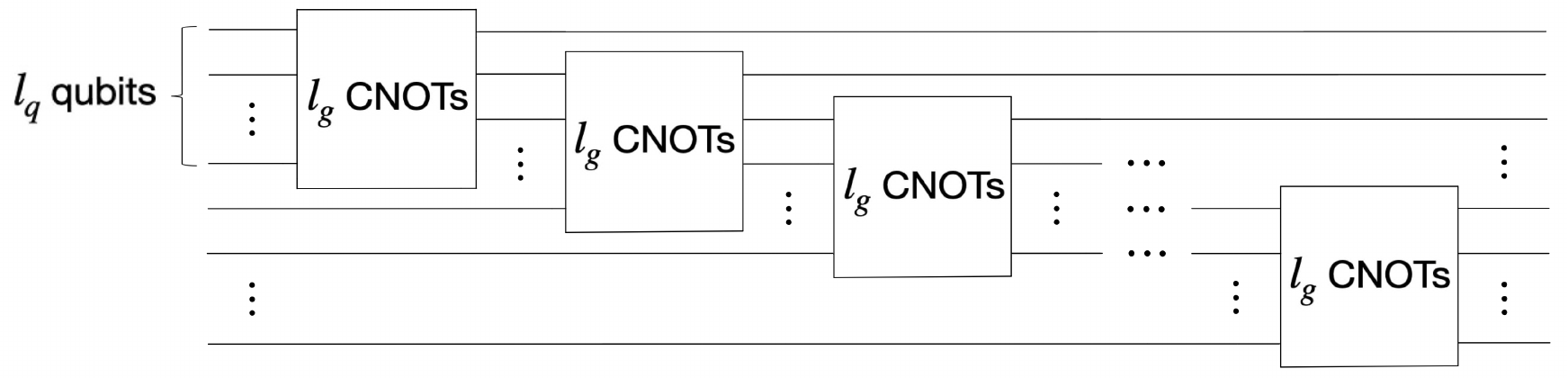}
    \caption{The construction of pseudo-realistic quantum circuits.}
    \label{fig:patt_cir}
\end{figure}

\begin{figure}
    \centering
    \includegraphics[width=0.4\textwidth]{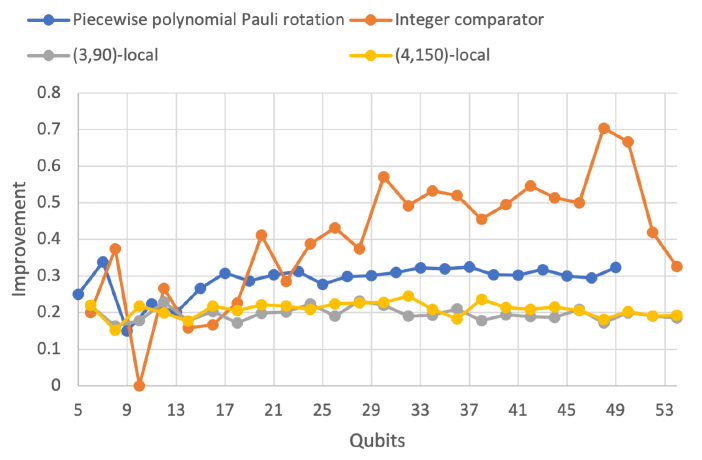}
    \caption{The experimental results of real circuits on Google Sycamore.}
    \label{fig:real_c}
\end{figure}

In the above subsection, we know that for some type of circuits, ADAC maintains a great improvement. We also
observed that ADAC maintains a notable advantage over large AGs for specific realistic circuits such as integer comparators and piecewise polynomial Pauli rotations, as depicted in Fig.~\ref{fig:real_c}.
To further investigate the adaptability of our method, we generated a series of test circuits designed to mimic the structure of these dominant circuits. These circuits referred to as pseudo-realistic quantum circuits, were tested on hypothetical grid-like AGs.
The pseudo-realistic quantum circuits can be described by two features: the local qubit number $l_\text{q}$ and local gate number $l_\text{g}$, 
and are hence termed \pseudorandom{l_\text{q}}{l_\text{g}} circuits thereafter. 
Roughly speaking, a 
\pseudorandom{l_\text{q}}{l_\text{g}} circuit can be constructed by iteratively (1) choosing $l_\text{q}$ qubits, (2) consecutively adding $l_\text{g}$ CNOT on the chosen qubits, (3) choosing an unused qubits to replace one of $l_{\text{q}}$ qubits, and repeating (2) and (3) until all given qubits are used.
The structure of the generated circuit is in a block-by-block manner, illustrated in Fig.~\ref{fig:patt_cir}. Similar patterns are observed in realistic circuits, such as piecewise Chebyshev and linear amplitude circuits\footnote{https://docs.quantum.ibm.com/api/qiskit/circuit\_library}. More importantly, the matrix product state ansatzes \cite{javanmard2024matrix}  widely used in variational quantum algorithms also have a similar pattern.

\begin{figure}
    \centering
    \subfloat[]{\includegraphics[width=0.45\textwidth]{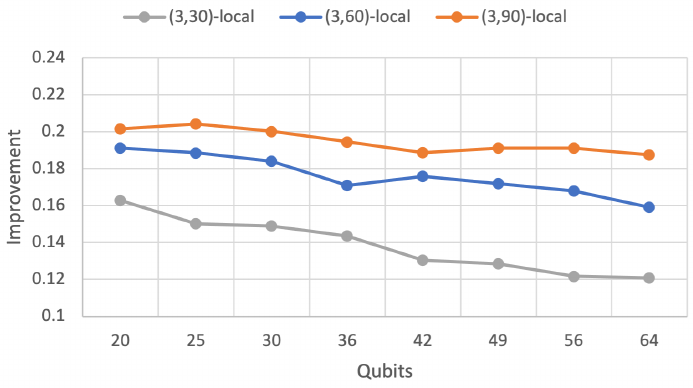}
    \label{subfig:pat3}
    }
    \hfill
    \subfloat[]{
    \includegraphics[width=0.45\textwidth]{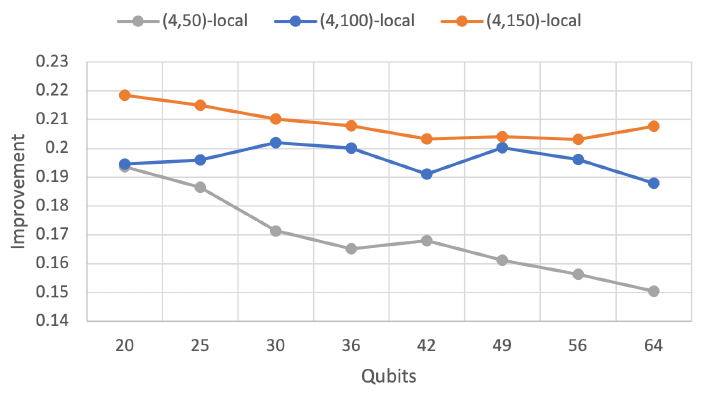}
    \label{subfig:pat4}
    }
    \caption{The experimental results of pseudo-realistic circuits on grid AGs.}
    \label{fig:pattern_r}
\end{figure}

As for the experiment, pseudo-realistic circuits under various parameters are tested. For each parameter setting, we generate 50 circuits using the described procedures and test them on, respectively, 4X5, 5X5, 5X6, 6X6, 6X7, 7X7, 7X8 and 8X8 grid AGs. The results in Fig.~\ref{fig:pattern_r} show that the improvement of ADAC compared to SABRE is consistently around 18\% when the local qubit number does not exceed 5. 
Besides that, we also evaluate realistic circuits, including piecewise polynomial Pauli rotation and integer comparator, which have similar local structures on the Google Sycamore architecture with 54 physical qubits. Surprisingly, the results in Fig.~\ref{fig:real_c} indicate a similar performance, with an improvement usually exceeding 20\%, compared to their hand-made counterparts.

In conclusion, the above results reveal an unambiguous scenario under which our ADAC algorithm can derive a high-quality solution regardless of the qubit number. Specifically, when the structure of the logical circuit resembles a combination of small blocks, the ADAC can be utilised to achieve better performance.

\section{Conclusion and discussions}

The foundation of our ADAC algorithm relies on subgraph isomorphism, which involves identifying connections between sub-circuits and the AG. This approach has limitations when dealing with sparse AGs and large-scale quantum circuits. This is because, under these conditions, the subcircuits contain too few CNOT gates, necessitating additional SWAP operations to establish connectivity between disconnected subcircuits. 

Future work will address these limitations. Improving the performance of our algorithm in such scenarios could involve exploring alternative optimisation strategies or enhancing the subgraph isomorphism approach. Additionally, advancements in quantum hardware design, such as increased qubit connectivity and fault tolerance, may help to overcome those challenges. Although the connectivity of most of IBM quantum hardwares are not dense, recently, the neutral atom quantum hardwares with long-range interaction~\cite{li23tcad_neutral_atom} achieve denser qubit connectivity. Our method may be extendable to such quantum hardwares.

In summary, we have introduced the ADAC (Adaptive Divide-and-Conquer) algorithm, a novel approach for solving the qubit mapping problem. By incorporating SWAP distance computation, adaptive circuit partitioning, and heuristic-based routing, our algorithm effectively reduces the SWAP overhead and, in return, improves the reliability for the execution of quantum circuits on quantum hardware. Experimental evaluations carried out on various NISQ devices and circuit benchmarks demonstrate the scalability, effectiveness, and performance of the ADAC algorithm. Our approach is able to outperform the state-of-the-art method in specific scenarios. These results highlight the potential of the ADAC algorithm to tackle the challenges during quantum compilation and expedite the deployment of practical applications for quantum computing.

\section*{Acknowledgments}

This work is supported by the Australian Research Council (DP220102059), the National Science Foundation of China (12071271) and the Innovation Program for Quantum Science and Technology (No.2021ZD0302901). We thank Yuan Feng and Pengcheng Zhu for helpful discussions and comments.

\bibliographystyle{IEEEtran}
\bibliography{qctref}
\end{document}